Title: **Impact bombardment on the regular satellites of Jupiter and Uranus during an episode of giant planet migration**

Authors: **E. W. Wong[1], R. Brasser[2] and S. C. Werner[3]**

Affiliations:
[1]Department of Physics, University of Hong Kong, Pokfulam Road, Hong Kong
[2]Earth Life Science Institute, Tokyo Institute of Technology, Meguro-ku, Tokyo 152-8550, Japan
[3]The Centre for Earth Evolution and Dynamics, Department of Geosciences, University of Oslo, 0315 Oslo, Norway

Abstract:
The intensity and effects of early impact bombardment on the major satellites of the giant planets during an episode of giant planet migration is still poorly known. We use a combination of dynamical N-body and Monte Carlo simulations to determine impact probabilities, impact velocities, and expected masses that collide with these satellites to determine the chronology of impacts during the migration. Volatile loss through bombardment is typically 20% for Miranda, a few percent for the larger Uranian satellites and negligible for the Galilean satellites. Due to its small size and the high impact velocity there is a >99% chance that Miranda suffered a catastrophic impact that shattered the satellite. Subsequent re-accretion from a circum-Uranian ring could account for its peculiar surface morphology and low density. The probability to destroy Ariel and Umbriel is 15% and 1% for Titania and Oberon. Approximately 90% of the mass in planetesimals that passes through the Jovian and Uranian satellite systems (about 4 $M_⊕$ and 2 $M_⊕$ respectively) does so in about 15 Myr. This extremely rapid and intense bombardment causes repeated local crustal melting on all satellites. The combination of these effects results in an entirely different impact chronology than that of the inner solar system. We conclude that the simple extrapolation of the lunar chronology to the outer solar system satellites is not correct. The tail end (after 25 Myr) of the chronology function has an e-folding time of 100 Myr at Jupiter, but follows a cumulative Weibull distribution at Uranus, making direct comparisons between the gas and ice giant planets difficult. Based on our results the surfaces of the Uranian satellites, Callisto, and possibly Ganymede, are all about the same age, and are roughly 150 Myr younger than the timing of the dynamical instability.

Keywords: cratering chronology; planet migration; impact flux; outer solar system; icy satellites

# 1. Introduction

## 1.1 Crater chronology and its challenges

The crater density on the lunar surface has long been used as a relative chronometer for the rest of the solar system. Dating planetary surfaces by crater densities was first proposed by Öpik (1960). Shortly thereafter Shoemaker et al. (1962) pioneered the linkage between crater densities and absolute ages, ultimately derived from radiometric dating of returned lunar samples. Consequently, the correlation of ages and crater densities for each lunar sampling site (Hartmann, 1970; Neukum et al., 1975) provided the age-calibrated lunar cratering rates and chronology models required to remotely date the solid surfaces of planets, moons and asteroids using crater counting alone.

The determination of absolute model ages for the Moon and other solid surface objects is based on the merging of the crater-production function and a suitable cratering chronology model. The former describes the expected crater size-frequency distribution observed from a particular geological unit of documented absolute age. It is assumed to be time-independent and, after proper scaling, is argued to be similar for all planetary bodies in the inner solar system (Hartmann 1973; Neukum & Wise, 1976; Neukum & Ivanov, 1994).

Historically, the two most commonly used chronologies describe a monotonic exponential decay in bombardment flux from about 4.1 Ga onwards. Extrapolation for times prior to about 4.1 Ga largely prohibits age determination before about 4.2 Ga (Werner 2014), even though there exist dated lunar samples as old as 4.42 Ga (Nemchin et al., 2009). Many challenges exist, however, in modelling this bombardment history.

Werner et al. (2014) set about obtaining a new calibrated cratering chronology for the Moon mostly based on returned samples, and Mars based only on martian meteorites. This allowed for simultaneous dating of lunar and martian surfaces as old as 4.4 Ga; the results of that study imply that the planetesimal-driven migration of the giant planets must have been triggered before about 4.25 to 4.3 Ga, otherwise the crater chronology curve would no longer be monotonic and would instead show an uptick around the time of giant planet migration (Morbidelli et al., 2012).

In theory it is possible to tie the bombardment of the outer solar system to that of the Moon, but it requires that the impact flux on the outer satellites and the Moon have the same decay profile (Neukum et al. 1998, 2006). This is unlikely to be the case because the dynamical evolution of planetesimals in the outer solar system proceeds much more rapidly than in the inner solar system due to the large mass difference between the giant and terrestrial planets. The methods that have been applied in the inner solar system are equally suited to the cratering chronology of the outer solar system (Neukum et al. 1998, 2006). As of yet, there are no recognised samples in our meteorite inventory from outer solar system. An absolute chronology in the outer solar system remains unfeasible at the present time, because we cannot link crater counts to radiometric ages. Thus, we require a different approach. This is demonstrated below.

**1.2. Objective and setting of the study**

It is generally accepted that leftover planetesimals from planetary accretion and the asteroid belt are the dominant contributions to the lunar and martian crater populations (Neukum & Ivanov 1994, Werner et al., 2002, Strom et al., 2005, Morbidelli et al., 2018). The contribution from comets, on the other hand, is still being debated (Rickman et al., 2017), and has been suggested to be minor for the Moon (Werner et al., 2002).

In contrast, comets are the dominant source of impacts on the surfaces of the Galilean and major Uranian satellites. The Jupiter Family Comets (JFCs) are the source of more than 90% of the craters on the Galilean satellites (Zahnle et al., 1998). The JFCs reside on short period, low inclination orbits, and their dynamics are controlled by Jupiter. The contribution from long period and nearly isotropic comets are expected to have contributed the remaining ~10% (Zahnle et al., 2003). Therefore, the source of impactors in the inner and outer solar system is not the same (Strom et al., 1981). This projectile source difference is expected to produce a distinct crater production function for the outer solar system as well as a crater chronology model description.

However, most studies of the outer solar system bodies' crater record indicates that it originates from a projectile population that is collisionally evolved, and that it closely resembles the inner solar system record (Neukum et al. 1998, 2006). The aim here is to construct a crater chronology function for the Galilean and Uranian satellites based on the dynamical evolution of the outer solar system during an episode of (late) giant planet migration. By 'late' we mean migration after the protosolar nebula has been dispersed, which is predicted to have occurred a few million years after solar system formation (Wang et al., 2017). There is ample evidence ranging from the orbital structure of the Kuiper Belt to the axial tilt of Saturn that such migration has taken place (Malhotra, 1993; Tsiganis et al., 2005; Nesvorný & Morbidelli, 2012; Brasser & Lee, 2015). In this study we do not consider the Saturnian satellite system because it appears its satellites formed in sequence (Charnoz et al., 2010).

The trigger for the instability is still not fully understood (Levison et al., 2011); dynamical models rely on either on the 2:1 mean-motion resonance crossing of the gas giants or the breaking of a quadruple resonance through angular momentum exchange with the planetesimal disc to cause an instability among the giant planets. Once the instability is triggered the giants become dynamically excited and scatter both each other and the planetesimal disc that remained beyond the outer ice giant throughout the solar system. These planetesimals are mostly ejected, with some ending in the Kuiper Belt, the Scattered Disc, the Oort cloud (Brasser & Morbidelli, 2013) or colliding with the gas giants and their satellites. A small fraction makes it into the inner solar system and thus they would not be a significant source of impacts on the rocky planets (Levison et al., 2001).

Assuming that the satellites existed before the giant planets began to migrate they are expected to experience an episode of extreme bombardment because of the enormous amount of mass that should have encounters with the giant planets deep enough to cross the orbits of the regular satellites. As such, these satellites could temporarily occupy the most geologically intense impact environment in the whole solar system. Intense bombardment within a short time period could release enough kinetic energy to potentially melt the satellites' crusts. This saturation and melting can significantly offset the surface ages from their formation ages by an amount comparable to the time difference between their formation and giant planet migration. The bombardment is likely to also cause volatile loss (Nimmo & Korycansky, 2012) and even complete satellite distruption (Movshovitz et al., 2015). Here we perform the first end-to-end calculation of the bombardment of the satellites during an episode of giant planet migration and explore its implications.

## 2. Methods

This study consists of three components. First, we establish the probability of planetesimals encountering the giant planets closer than 40 planetary radii, which is farther than all the regular satellites of the giant planets (apart from Iapetus, which may not be a regular satellite). We also determine the impact probability and rate onto the giant planets, as well as the rate at which planetesimals are eliminated from the system. Second, we re-enact planetesimal encounters with the giant planets closer than 40 planetary radii to determine the impact probability and average impact speed of planetesimals with the major Jovian and Uranian satellites. Third, we use Monte Carlo impact simulations to determine the amount of mass accreted by each satellite. All dynamical simulations were performed using the Gravitational ENcounters with Gpu Acceleration (GENGA) (Grimm & Stadel, 2014). The Monte Carlo (MC) simulations are based on the studies by Brasser et al. (2016) and Brasser & Mojzsis (2017). The GENGA simulations were run on the Earth Life Science Institute's GPU cluster equipped with Nvidia Tesla K-20 cards. The Monte Carlo simulations were run on R. Brasser's personal HTCondor[1] cluster.

---

1 HTCondor is scheduling software that allows High Throughput Computing (HTC) on large collections of distributively owned computing resources. http://www.cs.wisc.edu/htcondor

## 2.1. Dynamical simulations of giant planet migration

The initial conditions for the outer solar system dynamical simulations are similar to the loose five-planet model of Brasser & Lee (2015) based on Nesvorný & Morbidelli (2012). Initially the system consists of five giant planets, the two gas giants and three ice giants, in a 3:2 3:2 2:1 3:2 resonant configuration. The planets were surrounded by a disc of 30000 equal-mass planetesimals. We experimented with the initial position of Jupiter, the total disc mass and the outer edge of the disc to establish which combination of these parameters yielded the highest probability to reproduce the current system. Based on test simulations with fewer planetesimals we concluded that placing Jupiter initially at 5.6 AU, having the disc's outer edge be at 27 AU and the total disc mass of 18 Earth masses produced an outer solar system analogue that best matches the current planets.

We chose to use the five-planet model because it results in a higher chance of reproducing the current solar system (Nesvorný & Morbidelli, 2012; Brasser & Lee, 2015). This additional large body with a mass comparable to that of Uranus and Neptune will eventually be ejected from the solar system by Jupiter. The initial orbital eccentricities and inclinations of the giant planets and planetesimals were very small. The giant planets are in a resonant configuration but for the planetesimals the three remaining angles (longitude of the ascending node, argument of perihelion and mean anomaly) initially have uniformly random values between 0º and 360º.

Simulations were run for 1 Gyr with a time step of 146.1 days; during testing we also ran additional simulations with 5000 planetesimals for 200 Myr. Bodies were removed either through collision (perfect accretion was assumed) or when they were farther than 3000 AU or closer than 1.7 AU to the Sun. We further tasked GENGA to output the position and velocity vectors of every planetesimal that came within 40 radii of the giant planets. These vectors are then used for re-enacting the encounters during the second stage of this project. We ran a total of 64 simulations.

## 2.2 Re-enacting the close encounters

In order to determine the impact probability of planetesimals with the Jovian and Uranian satellites we need to re-enact their encounters with the giant planets. However, due to the finite amount of computing resources available and the minuscule impact probabilities with the satellites it is necessary to re-enact many more encounters than are obtained during the giant planet migration simulations. To do so we sampled the velocity and position distribution of the planetesimals as they penetrated the 40 planetary radii sphere. We established that the encounters are isotropic, agreeing with Zahnle et al. (1998), and we used that to construct the initial position and velocity vectors of the fictitious planetesimals. The magnitude of the velocity vector was sampled from the velocity distribution of planetesimals that passed through the sphere. We created 150 000 massless planetesimals in this manner per simulation; the maximum number of planetesimals was limited by GPU memory. All planetesimals passed by the satellites at once.

For the purpose of our calculations, we assumed that the satellites reside on their present-day orbits. The initial positions and velocities of the Jovian and Uranian satellites were obtained from the JPL Horizons[2] website. The time step was 0.01 days and the maximum simulation time was just 8 days, by which time most of the planetesimals had left the sphere. For those few planetesimals that impacted a satellite their impact velocities were recorded. We calculated the collision probability of a planetesimal with the satellites as the product of the probability to pass through the sphere around the planet and dividing the number of impacts on each satellite by the total number of planetesimals, i.e. $P_{imp} = (N_{imp}/N_{total})P_{enc}$, where $N_{imp}$ is the number of impacts, $N_{tot}$ is the total number of flyby planetesimals and $P_{enc}$ is the probability of a planetesimal entering the sphere around the planet. The total number of re-enacted flybys was 51 million for Jupiter and 91 million for Uranus. This higher number was necessary to obtain a reasonably accurate impact probability for Miranda for which we only recorded 24 impacts; the number of recorded impacts for the other satellites is of the order of 100.

**2.3. Monte Carlo impact simulations**

The Monte Carlo impact simulations undertaken for this study are discussed in detail in Brasser et al. (2016) and Brasser & Mojzsis (2017).

Planetesimals with diameters that range from 1 km to 2000 km (the diameter of Pluto) were randomly generated according to a pre-determined size-frequency distribution. We employed the simplest size-frequency distribution that is observed in the dynamically hot Kuiper belt and the Jupiter Trojan asteroids, with a change in slope at a diameter of ca. 60 km and cumulative slope indices of 4.8 and 1.9 and the high and low end, respectively (Fraser & Kavelaars, 2009). We also ran Monte Carlo simulations where the break occurs at ~140 km and the slopes are 4.35 and 1.8 at the high and low end, respectively (Fraser et al., 2014). We assume that this size-frequency distribution is the steady state outcome of collisional evolution and thus static with time. A more complicated choice based on the crater size-frequency distribution on Iapetus was also considered (Charnoz et al., 2009). These variations in the underlying size-frequency distribution of the planetesimals cause at most a factor of two in the crater densities of the satellites. A more detailed discussion of the effect of the size-frequency distribution of the planetesimals on the cratering statistics is reserved for future work.

Our chosen size-frequency distribution exhibits a knee at diameters of 60-140 km so that most of the mass resides in objects of approximately this size. We thus expect that the ratio of accreted masses between the satellites is approximately equal to their gravitational cross sections when accounting for the non-uniform pericentre distribution of the planetesimals as they fly past the planets. Due to the steep slope at the large end the total mass is finite and the number of impacts can, in principle, be calculated analytically. Nevertheless we prefer to validate our analytical estimates with these Monte Carlo simulations.

---

2   https://ssd.jpl.nasa.gov/horizons.cgi

We assumed all the planetesimals have a bulk density of 1400 kg m$^{-3}$, which is in between the typical density of a small comet (400-1000 kg m$^{-3}$) the Uranian and Saturnian satellites (1000-1700 kg m$^{-3}$), and Pluto (2000 kg m$^{-3}$). The smallest planetesimals have a diameter of 1 km. The effects of the bombardment can be scaled with assumed planetesimal density (see below).

For each planetesimal that was generated we compute a set of 4 (Jupiter) or 5 (Uranus) random numbers uniformly between 0 and 1; each of these is tied to a different satellite in increasing orbital position from the planet, i.e. random number #1 is tied to Io and #4 to Callisto. The purpose of each of these random numbers is to determine whether or not the planetesimal will collide with a satellite: if any one of these random numbers is lower than the impact probability with the corresponding satellite we assume that the planetesimal strikes the surface of said satellite and generate the next planetesimal. If none of the 4, or 5, random numbers are lower than the impact probabilities with the satellites then we assume no impact occurred and we generate the next planetesimal. We sum both the total mass of all the generated planetesimals and also how much mass impacts the surface of each satellite. The simulation continues to generate new planetesimals until their combined total mass reaches 18 Earth masses. The total number of planetesimals created in these simulations was close to 3 trillion. The first set of Monte Carlo simulations were applied to the Jovian satellites and the second to the Uranian system.

Following Brasser et al. (2016) we further keep track of how many basins (defined as craters with diameters > 300 km) as well as craters with diameter greater than 20 km are formed on the satellites. This will be a purely theoretical prediction since the actual number of small objects in the outer solar system is still unknown. The diameter of the final craters depends on the impact velocity, impact angle and material properties of impactors and target satellites. For simplicity we scaled the expected planetesimal diameters to those required for basin and 20-km crater formation on the Moon. From Brasser et al. (2016) the typical impact speed with the Moon is $v_i$ =16 km/s and from simple crater scaling laws the diameter of the planetesimal should exceed $D_{(>20km)}$=1.1 km and $D_{(basin)}$=23 km for excavating a 20-km crater and a basin, respectively. In this manner we were able to compute a rough estimate of $D_{imp(>20km)}$ and $D_{imp(basin)}$ for all nine satellites as $D_{imp} = (\rho_{imp}/\rho_{sat})^{-0.432} (v_{imp}/16 \text{ km s}^{-1})^{-0.568} (g_{sat}/1.625 \text{ m s}^{-2})^{0.284}$. For all satellites apart from Miranda $D_{imp\ (>20km)}$ lies between 0.98 km and 1.3 km, and $D_{imp(basin)}$ ranges from 18 km to 26 km. For simplicity we set the former to 1 km and for Miranda, for which $D_{imp\ (>20km)}$ is lower, we scale the expected number of craters accordingly; the threshold diameters of the basin-creating planetesimals are computed on a per-satellite basis

The predicted ratio between projectiles that create 20 km craters and those that form basins is of the order of 300-400 depending on the exact satellite; for the near-Earth objects the observed value could be up to 1000 (Neukum et al., 2001) while for the asteroid belt the observed value is near 200 (Ryan et al., 2015).

## 3. Results
In this section we describe the results of our numerical experiments. Each will be discussed in detail.

### 3.1. Giant planet migration
We ran a total of 32 giant planet migration simulations with GENGA for 1 Gyr. Of these, only two (~6%) successfully reproduced the current structure of the outer solar system based on the criteria of Brasser & Lee (2015), although we did not require that the difference in the longitudes of perihelion $\Delta\varpi=\varpi_J-\varpi_S$ of the gas giants circulates. In the simulation that best matches the current configuration of the giant planets, the innermost ice giant was ejected after ~6 Myr and the final semi-major axes of the gas giants are 5.20, 9.73, 20.38 and 30.05 AU. Their eccentricities are 0.04, 0.12, 0.07 and 0.04, respectively, and all inclinations are below 2.5º. Saturn's eccentricity is too high, which is a common problem with these simulations (see Figure 1). Neptune's eccentricity is also higher than is currently observed, but it is possible that this will damp over the next 3 Gyr due to gravitational interactions with remaining planetesimals. We plot the evolution of the semi-major axes, perihelia and aphelia of our best case for all planets in Figure 2.

The average impact probabilities of planetesimals onto Jupiter and Uranus are 0.76% and 0.59% respectively. Our $P_{Uranus}$ compared to $P_{Jupiter}$ is 0.78, which is a factor of 3 higher than that reported by Zahnle et al. (2003) (0.25). Thus the predicted impact probability on the Uranian satellites from our simulations should be a factor of 3 higher too. The probability to enter the sphere is approximately 40 times greater than impacting the planet.

The large increase in our $P_{Uranus}$ compared to that of Zahnle et al. (2003) warrants an explanation. The results of Zahnle et al. (2003) are based on the dynamical simulations of Levison & Duncan (1997), which were run in the current solar system environment. Levison et al. (2000) re-analyses the simulation output from Levison & Duncan (1997) to report direct impact probabilities of 0.9% and 0.5% with Jupiter and Uranus respectively, i.e. $P_{Uranus}/P_{Jupiter}$=0.55. However, the authors worry that the low number (10) of impacts on Uranus and Neptune causes inaccuracies and therefore also apply

Öpik's (1951) method to obtain an independent estimate of the impact probabilities. Using Öpik's method Levison et al. (2000) finds $P_{Uranus}/P_{Jupiter}=0.25$, which was subsequently adopted by Zahnle et al. (2003). It is not clear to us why Levison et al. (2000) obtained such a large difference between Öpik's method and counting direct impacts.

This still leaves our $P_{Uranus}$ some 50% greater than that of Levison et al. (2000). It is possible that this difference is caused by our simulations pertaining to an environment wherein the ice giants are migrating and are surrounded by a compact planetesimal disk. However, the only difference between these environments is that in our simulations the surface density of planetesimals as they cross the orbit of Uranus is much higher than in the current solar system, especially in the beginning. The same is true for Jupiter, and yet our impact probability with Jupiter is only slightly lower than that of Levison et al. (2000). Since the orbital and encounter velocity distributions in our simulations and those of Levison & Duncan (1997) are similar we conclude that the difference between our results and those of Levison et al. (2000) must be the result of numerical resolution: Levison et al. (2000) worked with only 1300 planetesimals and they recorded only 10 impacts onto Uranus while we use 30 000 and record hundreds of impacts.

### 3.2 Impact probability and impact speeds on the icy satellites

Generally, the ratio of impacts on the satellites versus the planet scales as as their gravitational cross sections, which is approximately $(R_{sat}/R_{pl})^2$ because of negligible satellite focusing; here $R_{sat}$ is the satellite radius and $R_{pl}$ the planet radius. However, the number of planetesimals that cross the orbit of a given satellite versus those that strike Jupiter or Uranus is approximately given by $a_{sat}/R_{pl}$ (Zahnle et al., 1998) because the encounter velocity with the planet is generally low compared to its escape velocity (Zahnle et al., 1998), and thus the planetesimals are strongly gravitationally focused. This results in a cumulative pericentre distribution $N(<q) \propto q$ so that the total impact probability becomes $P_{sat}/P_{pl} = (R_{sat}/a_{sat})^2 (a_{sat}/R_{pl})$ (Zahnle et al., 2003) where $a_{sat}$ is the satellite orbital semi-major axis. The impact probabilities to strike the Jovian and Uranian satellites and average impact velocities are listed in Table 1. Both the impact probability and impact speed agree to within a factor of a few with those of Zahnle et al. (2003). Uncertainties in the impact probabilities are at most 20%. Figure 3 shows the cumulative distribution of the impact velocity for all the satellites. Even though we assumed perfect accretion for the Monte Carlo simulations, in reality impact speeds are so high (much higher than the escape velocity of the satellites) that collisions are generally erosive rather than accretionary (Leinhardt & Stewart, 2012). The only difference are rare hit-and-run cases wherein a substantial amount of impactor material is torn off during the collision, which is then expected to accrete onto the target; these outcomes do not depend on the size of the impactors. The case of Miranda is the most extreme, where the impact speed is typically ~60 times greater than the escape speed of the satellite.

The Monte Carlo simulations record only the amount of mass that strikes the surfaces of the satellites, and not the amount that is actually accreted, which is a complicated function of impact parameter and impact speed. Indeed, the impact speeds are so high that impact ejecta from the satellites will be re-accreted to form secondary craters, or there could possibly even be inter-satellite debris exchange (sesquinary cratering). The outcomes of these more in-depth studies are left for future work.

| Satellite | Collision probability [$\times 10^{-7}$] | Average collision velocity [km/s] | Collided mass [wt %] | Collided mass [$\times 10^{19}$ kg] | Impact probability scaled to planet [$\times 10^{-5}$] |
|---|---|---|---|---|---|
| **Miranda** | 0.60 | 11.18 | 9.8 | 0.65 | 1.03 |
| **Ariel** | 1.68 | 10.72 | 1.34 | 1.81 | 2.86 |
| **Umbriel** | 1.45 | 9.97 | 1.33 | 1.56 | 2.48 |
| **Titania** | 2.15 | 8.68 | 0.66 | 2.32 | 3.68 |
| **Oberon** | 1.92 | 8.50 | 0.67 | 2.07 | 3.29 |
| **Io** | 4.93 | 33.03 | 0.060 | 5.32 | 6.49 |
| **Europa** | 3.04 | 26.56 | 0.068 | 3.28 | 4.01 |
| **Ganymede** | 5.05 | 23.23 | 0.037 | 5.45 | 6.65 |
| **Callisto** | 3.41 | 19.75 | 0.030 | 3.68 | 4.49 |

Table 1: Impact probability of a disc planetesimal with the major Jovian and Uranian satellites, average impact speed upon impact and total mass striking the satellites. The last column can be directly compared to Zahnle et al. (2003). Total mass accreted assuming that all collisions were accretionary.

### 3.3 Crater and basin densities

We used the output from the Monte Carlo simulations to calculate the total density per square kilometre of craters with diameters greater than 20 km, N(D>20), and for basins. The measured values of N(D>20) on Umbriel, Titania and Oberon are $0.9 \times 10^{-3}$, $1.0 \times 10^{-3}$ and $0.3 \times 10^{-3}$ km$^{-2}$ respectively (Plescia, 1987).

| Satellite | N(D>20) [$\times 10^{-3}$ km$^{-2}$] | N(Basin) [$\times 10^{-6}$ km$^{-2}$] |
|---|---|---|
| **Miranda** | 224 | 1790 |
| **Ariel** | 58.1 | 189 |
| **Umbriel** | 79.4 | 281 |
| **Titania** | 65.6 | 129 |
| **Oberon** | 70.7 | 156 |
| **Io** | 35.2 | 138 |
| **Europa** | 29.6 | 112 |
| **Ganymede** | 17.3 | 56.2 |
| **Callisto** | 13.9 | 41.8 |

Table 2: Expected crater densities on the Jovian and Uranian satellites assuming there was no resurfacing.

We assume that the moons have formed significantly earlier than this episode of giant planet migration because their formation must have coincided with that of giant planets (e.g. Canup & Ward, 2006), which all formed within the first few million years of the solar system (Wang et al., 2017). The calculated crater densities indicate that the surfaces of the satellites are very likely to have been completely reworked: the reported crater densities in Table 2 well exceed crater saturation levels. As such, the surface ages should be close to the time the bombardment caused by giant planet migration is most intense, but are unlikely to be much older.

In addition to resurfacing through cratering, the satellite's surfaces can also be modified through melting caused by large impacts. Numerical impact simulations by Barr & Canup (2010) show that the region shocked to >50% water ice melt is approximately spherical with radius $\chi r_i$, which is centred at a depth $\xi r_i$; here $\chi = 5.06(v_i/15 \text{ km s}^{-1})^{0.6}$ and $\xi = 2.85(v_i/15 \text{ km s}^{-1})^{0.47}$ and $r_i$ is the radius of the impactor. An approximation to the cumulative melt volume after $n$ large impacts is then $(4/3)\pi n <\chi r_i>^3$, where $<\chi r_i>$ is the average value integrated over the impactor size-frequency distribution. It is difficult to say how important crustal melting has been in the history of these satellites. Large impactors, with D>100 km or so, are theoretically capable of causing considerable local melt pools at depths of some 10 km and possibly deeper. This water melt mixes with colder clasts, cools and freezes. The satellites are bombarded with hundreds of impactors of this size and thus the expectation is that the surfaces of these satellites should be covered with frozen lakes rather than with craters. Yet this is not observed. Without detailed geometric and thermal modelling it is difficult to ascertain what the role of melting is in their geological history and we prefer to leave this for future work.

### 3.4 Volatile loss, Miranda and the case of catastrophic disruption

From Tables 1 and 2 it is clear that the Uranian satellites see a much more intense bombardment relative to their sizes than the Galilean satellites do, even though the latter are hit harder (Table 1). This bombardment is expected to cause a loss of volatiles on some of the satellites due to vaporisation upon impact.

Our results indicate that the mass striking the Uranian satellites is about a factor of 5 lower than that suggested by Nimmo & Korycansky (2012). We attribute this difference to Nimmo & Korycansky (2012) adopting the assumption of Barr & Canup (2010) that Callisto was struck by $3\times10^{20}$ kg of material and then using the impact probabilities listed in Zahnle et al. (2003) to scale the mass accreted by all the other icy satellites accordingly. Their assumed mass that impacted Callisto is a factor 8 greater than ours, and combined with differences in impact probabilities can account for the factor 5 difference between our results and theirs. The assumption of Barr & Canup (2010) that Callisto was struck by $3\times10^{20}$ kg of material coming from a 20 $M_\oplus$ disk requires an impact probability with Callisto of $2.5\times10^{-6}$, almost an order of magnitude higher than listed here and in Zahnle et al. (2003).

Nimmo & Korycansky (2012) predict that Miranda lost close to all of its volatiles, while our results suggest that Miranda only lost ~20% of its volatiles due to impacts (though see below). Its density suggests it contains 60% water ice by mass, so that the predicted initial value may have been 80%, much higher than the other satellites. For the remaining Uranian satellites the volatile loss amounts to at most a few percent. The result of Nimmo & Korycansky (2012) would predict a density of Miranda that is higher than that of the other satellites due to its extreme predicted volatile loss, and that the density of satellites should decrease as a function of distance to the planet due to a higher retention of volatiles. Instead, in Figure 4 we show that the opposite appears to be true: Miranda's density is about 1200 kg m$^{-3}$, lower than that of the other satellites which range from 1400 (Umbriel) to 1700 kg m$^{-3}$ (Ariel, Titania and Oberon).

For the Jovian satellites, on the other hand, the predicted amount of volatile loss is insignificant. Thus we expect most of the Uranian and outer Galilean satellites to remain volatile rich.

It has been suggested that Miranda's peculiar geology is the result of a catastrophic disruption, i.e. the satellite was shattered into smaller fragments upon impact with a large planetesimal; these fragments are gravitationally unbound from each other and are expected to form a circum-Uranian ring from which the current Miranda is subsequently re-accreted. Given the enormous amount of mass that struck Miranda it is natural to think that such an event occurred. Disruption occurs when the kinetic energy of impact is about 5.5 times the gravitational binding energy (Leinhardt & Stewart, 2012; Movshovitz et al., 2016). The corresponding diameter of a planetesimal to disrupt the satellite is then

$$D_i = 2\left(\frac{33U}{4\pi\rho_i v_i^2}\right)^{1/3},$$

where U=3/5 $GM_{sat}^2/R_{sat}$ is the total gravitational binding energy of the satellite. With nominal parameters the diameter of a planetesimal that can disrupt Miranda is ~42 km during a head-on collision, and about 60 km for a collision at a 45° angle; the former is at the 99.9 percentile of objects that impact the satellites in the Monte Carlo simulations. We find that in all of our Monte Carlo simulations there is a 100% chance that Miranda suffered a catastrophic disruption because the largest object that strikes Miranda in each simulation is consistently greater than 42 km (and even 60 km) in diameter. These findings are consistent with Movshovitz et al. (2015). The average diameter of the largest object to strike Miranda is 165 km out of 48 simulations (ranging from 105 km to 461 km). With the size-frequency given in Section 2.3, there are 13 objects with diameter 42 km for each one with a diameter 165 km, so that the actual probability of destroying Miranda is >99%.

The diameter of objects capable of destroying the other Uranian satellites is greater than 250 km. For Ariel and Umbriel the chance of a disruptive impact is ~15% while for Titania and Oberon it is ~1%. Miranda's disruption and subsequent re-accretion can explain its lower density than the other satellites if the re-accretion results in a more porous interior.

The Galilean satellites cannot be disrupted unless they are struck by objects comparable to Pluto in size. The probability for such a disruption is about 0.05% because there were of the order of 1000 Pluto-sized objects in the disc (Levison et al., 2011) and the impact probability with the Galilean satellites is ~$5\times10^{-7}$.

### 3.5 Crater chronology

Since we have no samples we cannot directly calibrate an absolute chronology of the outer solar system's satellites. However, it has recently been suggested that all of the giant planets underwent planetesimal-driven migration after the dissipation of the protoplanetary nebula but before the Moon-forming event (Morbidelli et al., 2018), i.e. before about 4.5 Ga (Barboni et al., 2017). Since the satellites likely formed within the first few million years of the solar system, early giant planet migration before the Moon-forming event implies that the surfaces of these satellites could be older than the lunar highlands, whose age is estimated to be 4.4 Ga (Werner et al., 2014). We can construct a potential chronology curve by plotting the timeline of planetesimals as they enter the spheres around Jupiter and Uranus, which is subsequently normalised to the maximum value of N(D>20) for each satellite obtained from the Monte Carlo simulations.

An example is given in Figure 5, which shows the chronology curves for Callisto and Titania. It is clear from the figure that the bombardment is extremely intense and short lived: over 90% of the planetesimal encounters have ceased some 15 Myr after the onset of the migration. Approximating this episode by an exponential decay results in an e-folding time of just 7 Myr, which is much shorter than the 140 Myr to 200 Myr of the lunar and martian chronologies (Neukum et al., 2001; Werner et al., 2014). However, the last 10% of planetesimal impacts occur at a much lower rate, and a best-fit exponential has an e-folding time of about 100 Myr at Jupiter, which is comparable to, but still lower than, its inner solar system analogue. In other words planetesimals in the outer solar system are eliminated more quickly. At Uranus, however, the profile follows a complementary cumulative

Weibull distribution, which is given by $e^{-\left(\frac{t}{\tau}\right)^{\beta}}$, where $\tau$ is the e-folding time and $\beta$ is the stretching parameter. A best fit yields $\tau$=3.3 Myr and $\beta$=0.38. Thus the chronology function is different at both planets. The crater densities on Ariel, Umbriel and Oberon are comparable to that on Titania so a naive approximation for their surface ages is about 150 Myr younger than the timing of the dynamical instability. Application to the observed value of N(D>20)~0.15×10$^{-3}$ km$^{-2}$ on Calliso (Strom et al., 1981) yields a similar age (indicated by the orange lines in Figure 5).

## 4. Discussion

In this section we discuss the comparison of our results with those in the literature, our assumptions and list some implications.

During our flyby simulations we assumed that the planetesimals were massless and that they flew past the satellites all at once. While this is certainly unrealistic, it served our sole purpose of establishing the impact probability with the satellites. A more realistic study beyond the scope of this one would use massive planetesimals and re-enact them according to when they passed through the sphere around the planet. In this manner potential perturbations on the orbits of the satellites can be determined. For example, the probability of a planetesimal to cross Callisto's orbit is approximately 20%, which implies that close to 4 $M_⊕$ crossed Callisto's orbit while close to 2 $M_⊕$ crosses the orbit of Oberon. This high amount of mass in planetesimal flybys has the potential to alter the orbits of the satellites, especially their eccentricity and inclination. These excitations would be damped through subsequent tidal dissipation within the satellites.

Barr & Canup (2010) investigated the interior structural dichotomy between Callisto and Ganymede: the latter shows strong indication of being differentiated while the former does not. Barr & Canup (2010) suggest that the different interior evolution could be the result of (late) bombardment on the satellites and that the different interiors are the result of differential bombardment between the two satellites. They calculate that Callisto will not differentiate if it was struck by less than 4×10$^{20}$ kg of material, while Ganymede will differentiate if Callisto receives more than 10$^{20}$ kg. By multiplying the impact probability by the total disc mass we calculate the mass in planetesimals colliding with Callisto is about 4×10$^{19}$ kg, and Ganymede is struck by 6×10$^{19}$ kg. This amount of impacting mass is low enough to keep Callisto mostly undifferentiated but insufficient to cause the differentiation of Ganymede. From our impact probabilities the required disc mass delivering enough material to both satellites to cause the dichotomy would need to be between 45 $M_⊕$ to 180 $M_⊕$, both of which exceed the mass constrained by dynamical simulations (Nesvorný & Morbidelli, 2012) and the population of small bodies in the outer solar system (Brasser & Morbidelli, 2013) that we used here.

We have calculated a crater density on all of the Uranian satellites that is at least a factor of a hundred higher than the published values for Umbriel, Titania and Oberon. For determining our crater densities, we assume that all craters stay recorded on the surface and that there is no overprinting and no melting. It also assumes that all planetesimals down to 1 km are intact, which may not be the case: detailed analysis of the size-frequency distribution of Jupiter-family comets shows a deficit of objects with diameters below 2 km (Snodgrass et al., 2011). Redoing the crater analysis for larger craters that correspond to impactors with diameters greater than 2 km will not overcome this problem because at its core is the total mass colliding with the satellites, which is fixed. The only two explanations for the lower observed crater densities are that either the surfaces are in saturation, and were overprinted time after time, or through geological activity. Crater equilibrium, or saturation, occurs when the density of craters is $N_{eq}=0.9D^{-2}$ km$^{-2}$ (Gault, 1970). For 20 km craters $N_{eq}=2.3\times10^{-3}$ km$^{-2}$. However, in reality equilibrium is already reached when $N\sim(0.01\text{-}0.1)N_{eq}$, i.e. when $N\sim(0.02\text{-}0.2)\times10^{-3}$ km$^{-2}$ (Gault, 1970; Hartmann, 1984; Richardson, 2009). The outer three Uranian satellites are at or above this value (Plescia, 1987). The observed craters therefore only recorded the tail end of the bombardment.

One further caveat to our work is that the Uranian satellite system could either be younger or older than 4.5 Ga. Morbidelli et al. (2012) showed that a hypothetical giant impact on Uranus that could have altered the direction of its spin axis would have also destroyed a primordial satellite system; a subsequent new one formed from the collisional debris in the equatorial plane of Uranus. The timing of this event, however, is not discussed in that work, but the authors suggest that this event was more likely to have occurred during the very early stages of the solar system's history when giant impacts were more common, probably when the giant planets were still forming in the first few million years of the solar system's history. Here we have assumed that the major Uranian satellites pre-date the Moon-forming event, as they are predicted to have (Ward & Canup, 2006; Wang et al., 2017).

We close this section with a few words about the satellites of Saturn and Neptune. A recent theory for the formation of most of the midsized Saturnian satellites (and possibly those of Uranus and Neptune too) is that they are the result of a primordial massive ring (Charnoz et al. 2010), wherein the rings spread until the Roche radius. Any material pushed beyond the Roche radius will clump together to form a moon(let) that is quickly pushed outwards by angular momentum exchange with the ring and tidal dissipation in Saturn (Charnoz et al., 2011; cf. Ćuk et al., 2016). Therefore, it is highly likely that most of the midsized Saturnian satellites are younger than the 4.5 Ga that is usually assumed and the determination of their ages becomes much more complex. In theory we could still have applied our method to Iapetus, but this would require setting the sphere of influence to beyond 65 planetary radii (Iapetus is at 61 Saturn radii). The number of flybys that would need to be re-enacted to determine the bombardment on all the other satellites of Saturn as well as the Jovian and Uranian ones would run into the hundreds of millions, which is very challenging. For these reasons we decided not to investigate the chronology of the Saturn system at this time.

In the case of Neptune, it is expected that Triton has destroyed any primordial satellite system that may have existed (Ćuk & Gladman, 2005; Nogueira et al., 2011; Rufu & Canup, 2017). The timing of this event is unknown, and the chronology of the Neptune system – mainly Triton – is made uncertain by the dynamical evolution of the system.

## 5. Conclusions

We have investigated the effects of the bombardment on the major Jovian and Uranian satellites during an episode of (late) giant planet migration. We have used a combination of dynamical N-body and Monte Carlo impact simulations to obtain the impact probability, impact speed, expected mass that collided with these satellites and the chronology of impacts after the instability. We find that our impact probabilities on the satellites scaled to that of the giant planets follow a similar trend to those of Zahnle et al. (2003) but have up to factors of three differences; contrary to that work we find that the impact probability with Uranus is a factor of three higher than theirs, and therefore the mass flux passing through the Uranian satellite system is correspondingly higher (about 2 $M_\oplus$ in total). The result of such a bombardment is that all the satellites have had their surfaces reset and their crater populations are in saturation-equilibrium. This leads us to conclude that the oldest surface ages of the Uranian satellites, as well as Callisto and the most ancient parts of Ganymede, could all be the same, and are comparable to the time of the onset of migration. Related to the bombardment, estimates of total volatile loss are possible and is typically a few percent for the larger Uranian satellites and may be about 20% for Miranda; these values are lower than those of Nimmo & Korycansky (2012), but the relative values follow the same trend. Volatile loss on the Galileans is expected to be negligible.

Due to its small size and high impact speeds, there is a >99% chance that Miranda suffered a catastrophic impact that shattered the satellite. Subsequent re-accretion could account for its peculiar geology and low density.

We report that about $4\times10^{19}$ kg struck Callisto, which is low enough to not differentiate Callisto (Barr & Canup, 2010), but the amount colliding with Ganymede is insufficient to explain its differentiation.

Last, we find that about 90% of the mass that passes through the Jovian and Uranian satellite systems does so in about 15 Myr. This extremely rapid and intense bombardment has a cratering rate decay that is inconsistent with that of the inner solar system, so that a simple extrapolation of the lunar chronology to the outer solar system satellites may not be possible, or it requires that the onset of giant planet migration occurred before or at the same time as the Moon-forming event. Under that assumption, the calculated crater densities on the surfaces of the Uranian satellites and Callisto and the most ancient parts of Ganymede are comparable to their observed values about 150 Myr after the onset of giant planet migration.

**Acknowledgements**

EW thanks RB for his hospitality at the Earth Life Science Institute where all of this work was conducted. RB gratefully acknowledges financial support from the JSPS International Joint Research Collaboration Fund (17KK0089). SCW is grateful for financial assistance from the Research Council of Norway through the Centre of Excellence funding scheme, project number 223272 (Centre for Earth evolution and Dynamics).

**Bibliography**

Barr, A. C., Canup, R. M. 2010. Origin of the Ganymede-Callisto dichotomy by impacts during the late heavy bombardment. Nature Geoscience 3, 164-167.

Bottke, W.F., D. Vokrouhlický, D. Minton, D. Nesvorný, A. Morbidelli, R. Brasser, B. Simonson, H.F. Levison (2012) An Archaean heavy bombardment from a destabilized extension of the asteroid belt. Nature 485, 78-81.

Brasser, R., Morbidelli, A. 2013. Oort cloud and Scattered Disc formation during a late dynamical instability in the solar system. Icarus 225, 40-49.

Brasser, R., Lee, M. H. 2015. Tilting Saturn without Tilting Jupiter: Constraints on Giant Planet Migration. The Astronomical Journal 150, 157-179.

Brasser, R., Mojzsis, S. J., Werner, S. C., Matsumura, S., Ida, S. 2016. Late veneer and late accretion to the terrestrial planets. Earth and Planetary Science Letters 455, 85-93.

Brasser, R., Mojzsis, S. J. 2017. A colossal impact enriched Mars' mantle with noble metals. Geophysical Research Letters 44, 5978-5985.

Canup, R. M., Ward, W. R. 2006. A common mass scaling for satellite systems of gaseous planets. Nature 441, 834-839.

Charnoz, S., Morbidelli, A., Dones, L., Salmon, J. 2009. Did Saturn's rings form during the Late Heavy Bombardment? Icarus 199, 413-428.

Charnoz, S., Salmon, J., Crida, A. 2010. The recent formation of Saturn's moonlets from viscous spreading of the main rings. Nature 465, 752-754.

Charnoz, S., Crida, A., Castillo-Rogez, J. C., Lainey, V., Dones, L., Karatekin, Ö., Tobie, G., Mathis, S., Le Poncin-Lafitte, C., Salmon, J. 2011. Accretion of Saturn's mid-sized moons during the viscous spreading of young massive rings: Solving the paradox of silicate-poor rings versus silicate-rich moons. Icarus 216, 535-550.

Ćuk, M., Gladman, B. J. 2005. Constraints on the Orbital Evolution of Triton. The Astrophysical Journal 626, L113-L116.

Ćuk, M., Dones, L., Nesvorný, D. 2016. Dynamical Evidence for a Late Formation of Saturn's Moons. The Astrophysical Journal 820, 97.

Fraser, W. C., Kavelaars, J. J. 2009. The Size Distribution of Kuiper Belt Objects for D > 10 km. The Astronomical Journal 137, 72-82.

Fraser, W. C., Brown, M. E., Morbidelli, A., Parker, A., Batygin, K. 2014. The Absolute Magnitude Distribution of Kuiper Belt Objects. The Astrophysical Journal 782, 100.

Gault, D. E. 1970. Saturation and equilibrium conditions for impact cratering on the lunar surface: Criteria and implications. Radio Science 5, 273-291.

Grimm, S. L., Stadel, J. G. 2014. The GENGA Code: Gravitational Encounters in N-body Simulations with GPU Acceleration. The Astrophysical Journal 796, 23.

Hartmann, W. K. 1970. Preliminary note on lunar cratering rates and absolute time-scales. Icarus 12, 131-133.

Hartmann, W. K. 1973. Martian cratering 4: Mariner 9 initial analysis of cratering chronology. Journal of Geophysical Research 78, 4096–4116.

Levison, H. F., Duncan, M. J. 1997. From the Kuiper Belt to Jupiter-Family Comets: The Spatial Distribution of Ecliptic Comets. Icarus 127, 13-32.

Levison, H. F., Duncan, M. J., Zahnle, K., Holman, M., Dones, L. 2000. NOTE: Planetary Impact Rates from Ecliptic Comets. Icarus 143, 415-420.

Levison, H. F., Dones, L., Chapman, C. R., Stern, S. A., Duncan, M. J., Zahnle, K. 2001. Could the Lunar "Late Heavy Bombardment" Have Been Triggered by the Formation of Uranus and Neptune? Icarus 151, 286-306.

Levison, H. F., Morbidelli, A., Tsiganis, K., Nesvorný, D., Gomes, R. 2011. Late Orbital Instabilities in the Outer Planets Induced by Interaction with a Self-gravitating Planetesimal Disk. The Astronomical Journal 142, 152.

Malhotra, R. 1993. The origin of Pluto's peculiar orbit. Nature 365, 819-821.

Morbidelli, A., Tsiganis, K., Batygin, K., Crida, A., Gomes, R. 2012. Explaining why the uranian satellites have equatorial prograde orbits despite the large planetary obliquity. Icarus 219, 737-740.

Morbidelli, A., Nesvorny, D., Laurenz, V., Marchi, S., Rubie, D. C., Elkins-Tanton, L., Wieczorek, M., Jacobson, S. 2018. The timeline of the lunar bombardment: Revisited. Icarus 305, 262-276.

Movshovitz, N., Nimmo, F., Korycansky, D. G., Asphaug, E., Owen, J. M. 2015. Disruption and reaccretion of midsized moons during an outer solar system Late Heavy Bombardment. Geophysical Research Letters 42, 256-263.

Movshovitz, N., Nimmo, F., Korycansky, D. G., Asphaug, E., Owen, J. M. 2016. Impact disruption of gravity-dominated bodies: New simulation data and scaling. Icarus 275, 85-96.

Nemchin, A., Timms, N., Pidgeon, R., Geisler, T., Reddy, S., Meyer, C. 2009. Timing of crystallization of the lunar magma ocean constrained by the oldest zircon. Nature Geoscience 2, 133-136.

Nesvorný, D., Morbidelli, A. 2012. Statistical Study of the Early solar system's Instability with Four, Five, and Six Giant Planets. The Astronomical Journal 144, 117.

Neukum, G., Koenig, B., Arkani-Hamed, J. 1975. A study of lunar impact crater size-distributions. The Moon 12, 201-229.

Neukum, G., Wise, D. U. 1976. Mars – A standard crater curve and possible new time scale. Science 194, 1381–1387.

Neukum, G., Ivanov, B. A. 1994. Crater size distributions and impact probabilities on Earth from lunar, terrestrial-planet, and asteroid cratering data. In: *Hazards Due to Comets and Asteroids*, Gehrels T. (ed.), University of Arizona Press, Tucson, AZ, USA. pp 359–416.

Neukum, G., Wagner, R., Wolf, U., Ivanov, B. A., Head, J. W. III, Pappalardo, R. T., Klemaszewski, J. E., Greeley, R., Belton, M. J. S., and the Galileo SSI Team (1998). Cratering chronology in the Jovian system and derivation of absolute ages. Lunar Planet. Sci. Conf. XXIX, abstr. No. 1742.

Neukum, G., Ivanov, B. A., Hartmann, W. K. 2001. Cratering Records in the Inner solar system in Relation to the Lunar Reference System. Space Science Reviews 96, 55-86.

Neukum, G., Wagner, R., Wolf, U., Denk, T. (2006). The cratering record and cratering chronologies of the Saturnian satellites and the origin of impactors: results from Cassini ISS Data. Europ. Planet. Sci. Congr. 2006, abstr., p. 610.

Nimmo, F., Korycansky, D. G. 2012. Impact-driven ice loss in outer solar system satellites: Consequences for the Late Heavy Bombardment. Icarus 219, 508-510.

Nogueira, E., Brasser, R., Gomes, R. 2011. Reassessing the origin of Triton. Icarus 214, 113-130.

Öpik, E. J. 1951. Collision probability with the planets and the distribution of planetary matter. Proc. R. Irish Acad. Sect. A 54, 165-199.

Öpik, E. J. 1960. The Lunar Surface as an Impact Counter. Monthly Notices of the Royal Astronomical Society 120, 404-411.

Plescia, J. B. 1987. Cratering history of the Uranian satellites - Umbriel, Titania, and Oberon. Journal of Geophysical Research 92, 14918-14932.

Richardson, J. E. 2009. Cratering saturation and equilibrium: A new model looks at an old problem. Icarus 204, 697-715.

Rickman, H., Wisniowski, T., Gabryszewski, R., Wajer, P., Wojcikowski, K., Szutowicz, S., Valsecchi, G. B., Morbidelli, A. 2017. Cometary impact rates on the Moon and planets during the late heavy bombardment. Astronomy and Astrophysics 598, A67.

Rufu, R., Canup, R. M. 2017. Triton's Evolution with a Primordial Neptunian Satellite System. The Astronomical Journal 154, 208.

Ryan, E. L., Mizuno, D. R., Shenoy, S. S., Woodward, C. E., Carey, S. J., Noriega-Crespo, A., Kraemer, K. E., Price, S. D. 2015. The kilometer-sized Main Belt asteroid population revealed by Spitzer. Astronomy and Astrophysics 578, A42.

Shoemaker, E. M., Hackman, R. J., Eggleton, R. E. 1962. Interplanetary correlation of geologic time. Advances in Astronautical Science 8, 70-89.

Snodgrass, C., Fitzsimmons, A., Lowry, S. C., Weissman, P. 2011. The size distribution of Jupiter Family comet nuclei. Monthly Notices of the Royal Astronomical Society 414, 458-469.

Strom, R. G., Woronow, A., Gurnis, M. 1981. Crater populations on Ganymede and Callisto. Journal of Geophysical Research 86, 8659-8674.

Tsiganis, K., Gomes, R., Morbidelli, A., Levison, H. F. 2005. Origin of the orbital architecture of the giant planets of the solar system. Nature 435, 459-461.

Wang, H., Weiss, B. P., Bai, X.-N., Downey, B. G., Wang, J., Wang, J., Suavet, C., Fu, R. R.,

Zucolotto, M. E. 2017. Lifetime of the solar nebula constrained by meteorite paleomagnetism. Science 355, 623-627.

Werner, S. C. 2014. Moon, Mars, Mercury: Basin formation ages and implications for the maximum surface age and the migration of gaseous planets. Earth and Planetary Science Letters 400, 54–65.

Werner, S. C., Ody, A., Poulet, F. 2014. The Source Crater of Martian Shergottite Meteorites. Science 343, 1343-1346.

Zahnle, K., Dones, L., Levison, H. F. 1998. Cratering Rates on the Galilean Satellites. Icarus 136, 202-222.

Zahnle, K., Schenk, P., Levison, H., Dones, L. 2003. Cratering rates in the outer solar system. Icarus 163, 263-289.

# Figures

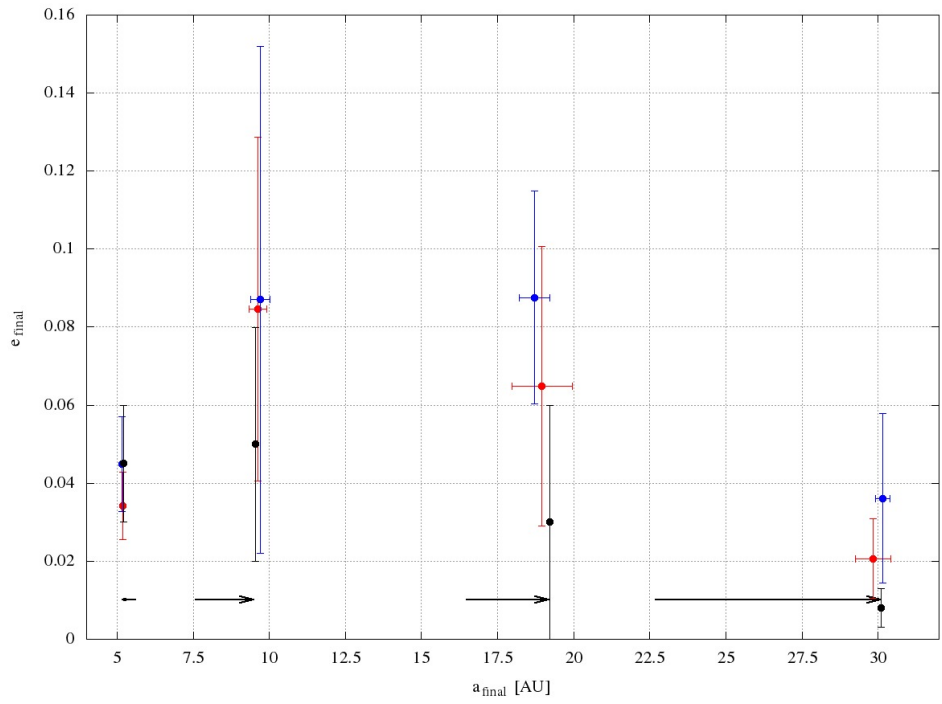

**Figure 1.** Migration of the gas giants for two different disc masses. Black bullets and error bars show the current mean eccentricities and their secular variation. Red dots are for a disk with a mass of 19 $M_\oplus$ and blue for 18 $M_\oplus$.

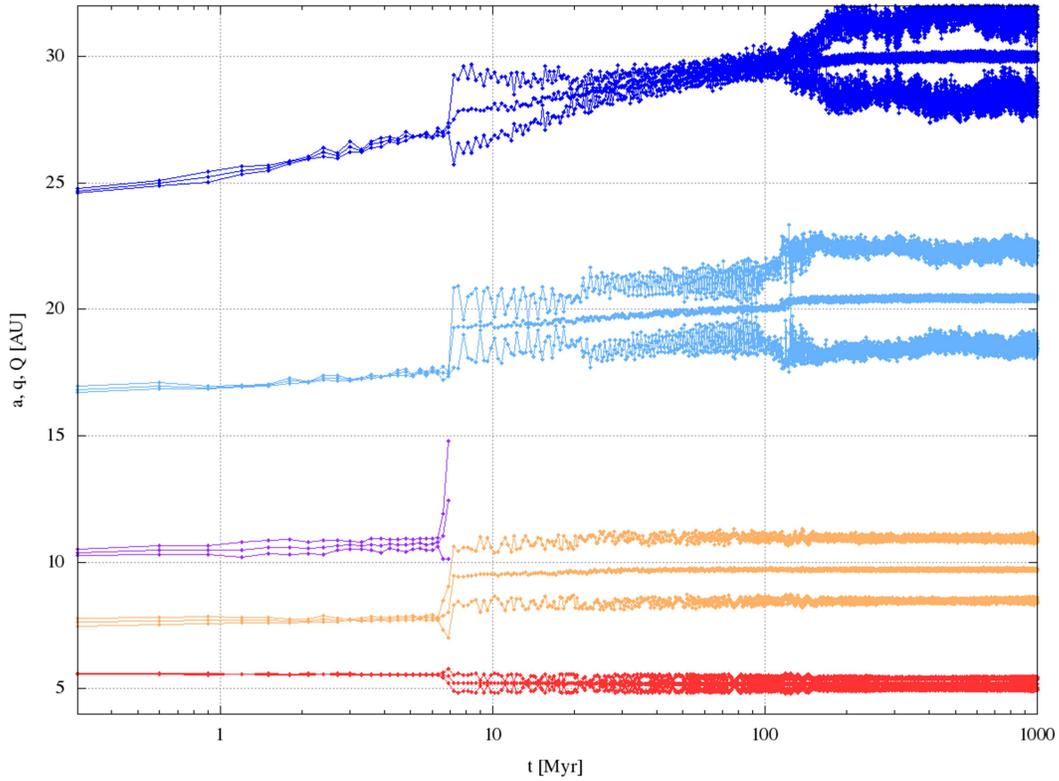

**Figure 2.** Evolution of the giant planets with time. We show the semi-major axis, perihelion and aphelion distances. The higher the eccentricity the greater the distance between the perihelion and aphelion. The sudden increase in the semi-major axis of Uranus and eccentricities of Uranus and Neptune are caused by the third ice giant temporarily residing on a highly eccentric orbit and having encounters with Uranus and Neptune.

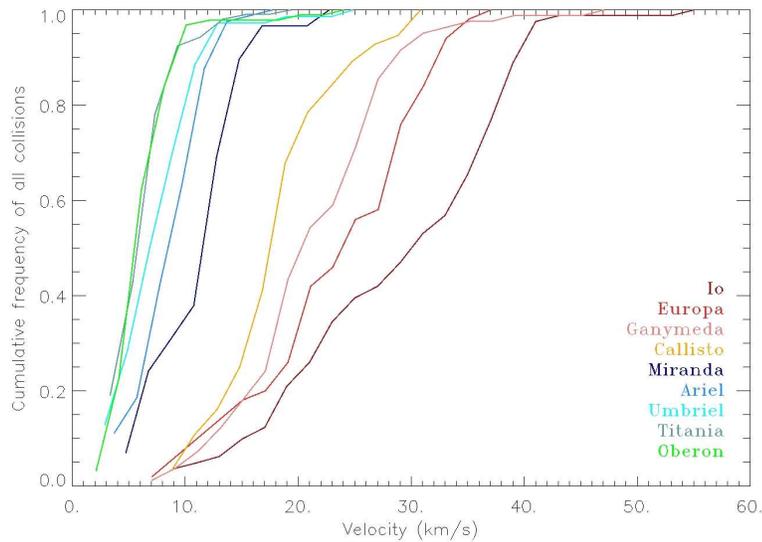

**Figure 3.** Cumulative impact speeds of planetesimals with the Uranian and Galilean satellites.

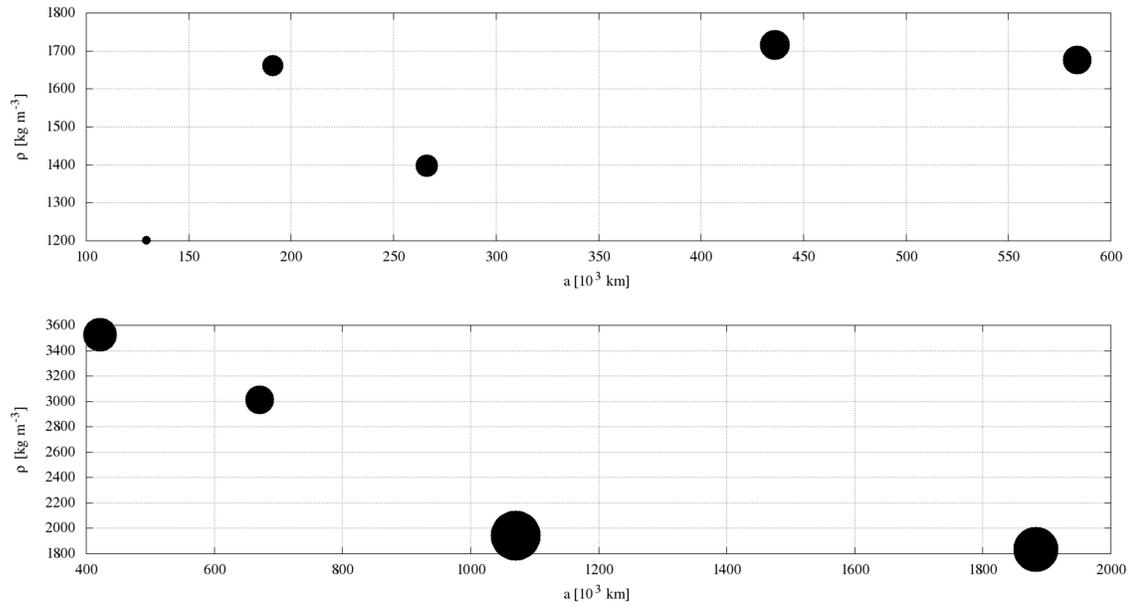

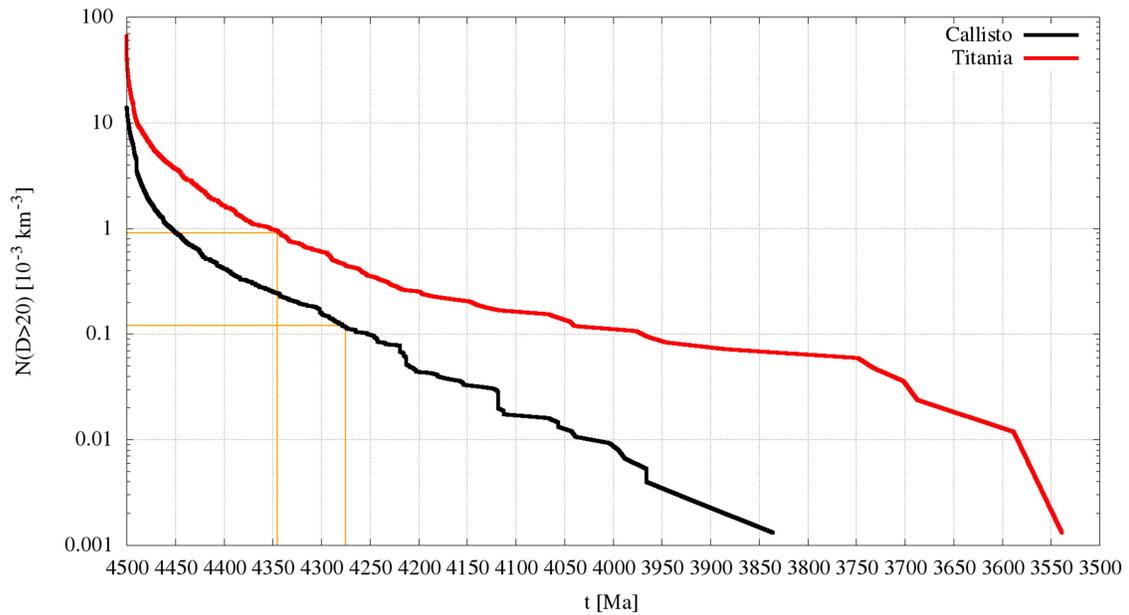

**Figure 4.** Bulk density versus semi-major axis for the Uranian (top) and Galilean (bottom) satellites. The loss of volatiles predicts a decrease in density as a function of semi-major axis, which appears to not be true for the Uranians.